\documentclass[12pt,a4paper]{article}

 \usepackage[dvips]{graphics}
 \usepackage{graphicx}
 \usepackage{afterpage}
 \usepackage{enumerate}
\begin{document}

 \title{\bf Formation of small-scale magnetic elements: surface mechanism
  }

 \author{\bf A.S. Gadun$^1$, V.A. Sheminova$^1$,  and S. K. Solanki$^2$ }
 \date{}

 \maketitle
 \thanks{}
\begin{center}
{$^!$Main Astronomical Observatory, National Academy of Sciences of Ukraine
\\ Zabolotnoho 27, 03689 Kyiv, Ukraine \\
\vspace{0.5cm}
 $^2$Max-Planck-Institut fur Aeronomie \\ Max-Planck Str., D-37191
Katlenbwg-Lindau }
\end{center}

 \begin{abstract}
We present the first results of a two-dimensional MHD simulation of the solar
magnetogranulation. The medium was assumed to be compressible, gravitationally
stratified, radiatively coupled, partially ionized, and turbulent. The simulated
magnetogranulation evolved over the course of two hours of hydrodynamic (solar)
time. A surface (magnetic plume-like) mechanism which forms thin magnetic elements
was found to operate during the process of granule fragmentation. The activity of
such a mechanism suggests that the magnetogranulation can concentrate and intensify
the global magnetic flux at the boundaries of convective cells and can also form
nearly vertical compact magnetic flux tubes by involving the weak horizontal
photospheric field, which may be, in general, of local (turbulent) nature.
\end{abstract}

\section{Introduction}

When small-scale magnetic structures were discovered in 1973 \cite{Dunn,Stenflo},
the investigators became aware of the fact that the major part of the magnetic flux
on the Sun, outside sunspots and pores, is concentrated in these thin elements
which are no more than several hundred kilometers in size but have magnetic fields
of several hundred millitesla. These features have been the object of much
observational and theoretical research (see the reviews by Solanki
\cite{Solanki90,Solanki93}, Sch\"{u}ssler \cite{Schussler90,Schussler98}, and
M\"{u}ller \cite{Muller}). Of special interest is a multidimensional simulation of
the interaction between the thermal granulation-scale convection and the magnetic
field; this simulation can be done by solving a system of equations of radiative
magnetohydrodynamics for an inhomogeneous, compressible, and gravitationally
stratified medium. With such time-dependent models the mechanisms of formation of
compact magnetic elements can be studied in detail, and the adequacy of the
simulation can be tested as well by comparing the calculated and observed Stokes
profile parameters of spectral lines.

The initial magnetic field intensity and configuration are important parameters in
such calculations. In most studies (e.g.,
\cite{Atroshchenko1,Atroshchenko2,Grossmann}) the initial field was assumed to be
of global nature, and it was therefore homogeneous, vertically oriented, and it
filled the entire calculation region. The function of the granulation in this case
is to concentrate the magnetic field near the boundaries of convective cells and
intensify the field. The simulation in \cite{Atroshchenko1,Atroshchenko2,Grossmann}
lasted only from five to nine minutes of hydrodynamic time, and the authors could
follow the effect of kinematic and superadibatic mechanisms on the formation of
magnetic structures only over the lifetime of those convective flows (``cells'')
which existed from the outset. The effect of granule fragmentation (diffusion) on
the evolution of magnetic structures could not be investigated.

A different initial field was taken by Brandt and Gadun \cite{Brandt}. It also was
homogeneous and vertically oriented, but it had the only horizontal scale (70~km),
and its initial intensity was 8.5~mT. The time of magnetogranulation simulation was
quite long (48 min). However, the granulation convection was assumed to be
quasi-stationary rather than nonstationary (two convective flows of the same scale
existed in the dimensional simulation region --- the symmetrical initial conditions
permitted them to coexist for an indefinitely long time). It was found,
nevertheless, that a magnetic configuration with an intensity of about 100~mT may
form with such a weak local priming field and a strong convective instability. The
topological pumping and the density gradient effect \cite{Vainshtein} ensure the
initial intensification of the magnetic field in deeper layers; the kinematic
mechanism concentrates the field lines in the vicinity of downflows. The
superadiabatic (thermal) effect is the dominant factor in the intensification of
the magnetic field; it grows when the magnetic field energy is close to the
equipartition level or exceeds it. In this case the magnetic field suppresses the
convective flows, and the magnetic configuration situated in the region of
downflows suddenly begins to descend.

So, we may state that all published attempts of the direct numerical simulation of
magnetoconvection ignore a fundamental feature of the solar thermal convection --
its nonstationarity. There are also some problems in the formulation of initial
conditions for the magnetic field.

In this study we present some results of further development of the two-dimensional
numerical simulation of magnetogranulation. Nonstationary magnetogranulation was
simulated over the course of two hours of hydrodynamic time with the use of a new
concept of initial magnetic field. We show that the fragmentation of granules in
the presence of horizontal photospheric fields results in the formation of new
small-scale magnetic structures. This mechanism is given the name surface
mechanism. We think it important for an understanding of magnetogranulation in
various objects to establish the role of this mechanism in the formation of flux
tubes, but here we only describe it, and a detailed analysis of the simulation
results will be published elsewhere.

Small-scale magnetic elements are usually called flux tubes
\cite{Solanki90,Solanki93}. Of course, they are not tubes in the two-dimensional
plane representation, but we arbitrarily reserve this traditional name for them.

\section{Two-dimensional radiative magnetohydrodynamics
}

The complete set of equations of radiative MHD for the conditions of compressible
gravitationally stratified turbulent medium had the following form:

\[
  \begin{array}{l}
  \vspace{0.2cm}
    \frac{\partial \rho}{\partial t}+\mathbf{\nabla}\rho \mathbf{V} =0\,,\\
\vspace{0.2cm}
    \frac{\partial \rho \mathbf{V}}{\partial t}= -\nabla \left[ \rho
\mathbf{VV} +
    \left(P+ \frac{B^2}{8\pi}\right)\mathbf{I} - \frac{\mathbf{BB}}{4\pi}+
    \mathbf{R}\right]-\rho \mathbf{g}\,,     \\
\vspace{0.2cm}
     \frac{\partial \rho U}{\partial t}= -\mathbf{\nabla} \left[ (\rho U+P+\mathbf{R})
     \mathbf{V}-\frac{\mathbf{B}}{4\pi}(\mathbf{VB})+
     \frac{D_m}{4\pi}[[\mathbf{\nabla},\mathbf{B}],\mathbf{B}]\right]+\rho q_D-Q_R-\rho g\upsilon _z\,,      \\
\vspace{0.2cm}
      \frac{\partial \mathbf{\mathbf{A}}}{\partial t}= [\mathbf{V},\mathbf{B}]-
      D_m [ \mathbf{\nabla},\mathbf{B}], ~~~~~~~~       \mathbf{B}=[\mathbf{\nabla}, \mathbf{A}].  \\

  \end{array}
\]

Here $\mathbf{A}$ is the vector potential, $\mathbf{I}$ is a unit tensor,
$\mathbf{R}$ is the Reynolds turbulence stress tensor, $\mathbf{g}(g)$ is the free
fall acceleration, $U = E + B^2/8\pi/\rho$ is the total specific energy ($E =
\upsilon^2{/}2 + e$), $e$ is the internal energy, $q_D$ is the kinetic energy
dissipation of the averaged motion at the subgrid level, $D_m$ is the magnetic
diffusion coefficient, $Q_R$ is the divergence of the radiation flux vector
(radiation heating/cooling); the spatial rectangular coordinates $x$ and $z$
describe the simulation region in the horizontal and vertical directions. We also
assume the magnetic induction $\mathbf{B}$ to be equal to the magnetic field
intensity $\mathbf{H}$, the magnetic permeability being equal to 1.

We used the ideal gas equation in our calculations; the radiation pressure and
possible changes in the electron density due to hydrogen ionization and double
ionization of 15 elements were taken into account. The contribution of H$^-$ and
H$_2$, H$_2^+$ molecules to the ionization equilibrium was taken into account at
$T\leq 6000$~K.

The set of equations and the solution of the transfer equation used to determine
$Q_R$ are described in detail in \cite{Brandt,Gadun95,Gadun96}. The radiation
effects were treated in our simulation in the "grey" approximation.

Free upper and lower boundary conditions were taken for velocities and
thermodynamic quantities, i.e., the free inflow and outflow of matter were
permitted: the velocity components were determined from the condition $\partial
V{/} \partial z = 0$, and the mean internal energy and mean density were taken from
the initial homogeneous model \cite{Gadun95,Gadun96}. Their fluctuation profiles at
the upper (lower) boundary were the same as in the layer situated below (above).
The density at the lower boundary was scaled to provide a constant sum of the gas,
radiation, and magnetic pressures at the horizontal level. The upper and lower
boundary conditions for the magnetic field were set, assuming that it was of global
nature: $B_x = 0$, $\partial B{/} \partial z = 0$. Periodic side conditions were
postulated.

The size of simulation region was $3920 \times 1820$~km, the spatial step was 35
km, the atmospheric layers extended over about 700~km. The internal structure of
thin magnetic configurations cannot be studied in detail with such a spatial step,
but it is sufficiently large for the evolution of these configurations to be
followed and the calculation costs to decrease.

We calculated two model sequences with a magnetic field and without field, with the
same initial conditions. The initial model was taken from the time series of HD
models \cite{Gadun99}, but the simulation region was diminished.

\begin{figure}
   \centering
   \includegraphics[width=8.cm]{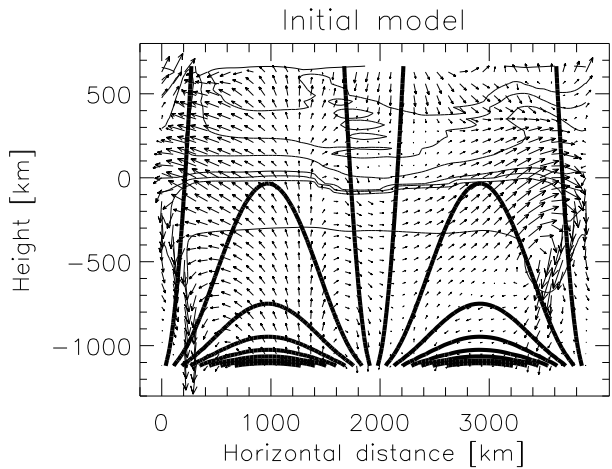}
 \hfill
\parbox[b]{7.cm}{ \vspace{0.0cm}
   \caption[]{
Initial model and initial magnetic field configuration in the MHD simulation. Thin
lines) isotherms for temperatures (from the top down): 4000, 5000, 6000, 7000,
8000, 9000, 10000, and 12000 K. Thick lines) magnetic field lines.
 } \label{Fig:Fig1}
}

\end{figure}

Figure~\ref{Fig:Fig1} shows the initial magnetic field configuration and the
initial model. This is a bipolar configuration with a field intensity decreasing
with height. The mean intensity $B$ was 5.4~mT throughout the calculation region.

After several similar calculation runs with different initial conditions for the
magnetic field (including a homogeneous vertically directed field) we concluded
that the configuration shown in Fig.~\ref{Fig:Fig1} is the most natural one
(uniformly distributed). It also provides the optimal conditions for the numerical
stability of the solution at the initial moment of simulation, since the velocity
field and the field of thermodynamic quantities, on the one hand, and the magnetic
field characteristics, on the other hand, turn out not to be self-consistent in the
initial model.

\section{Results
}

\begin{figure}
   \centering
   \includegraphics[width=7.cm]{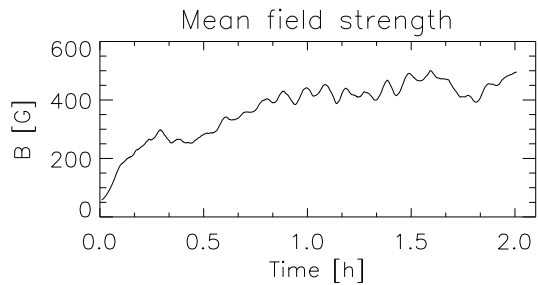}
 \hfill
\parbox[b]{8cm}{ \vspace{0.0cm}

   \caption[]{
Evolution of the mean magnetic field intensity in the simulation region.
 } \label{Fig:Fig2}
 }

\end{figure}

\begin{figure}[t]
   \centering
   \includegraphics[width=13.cm]{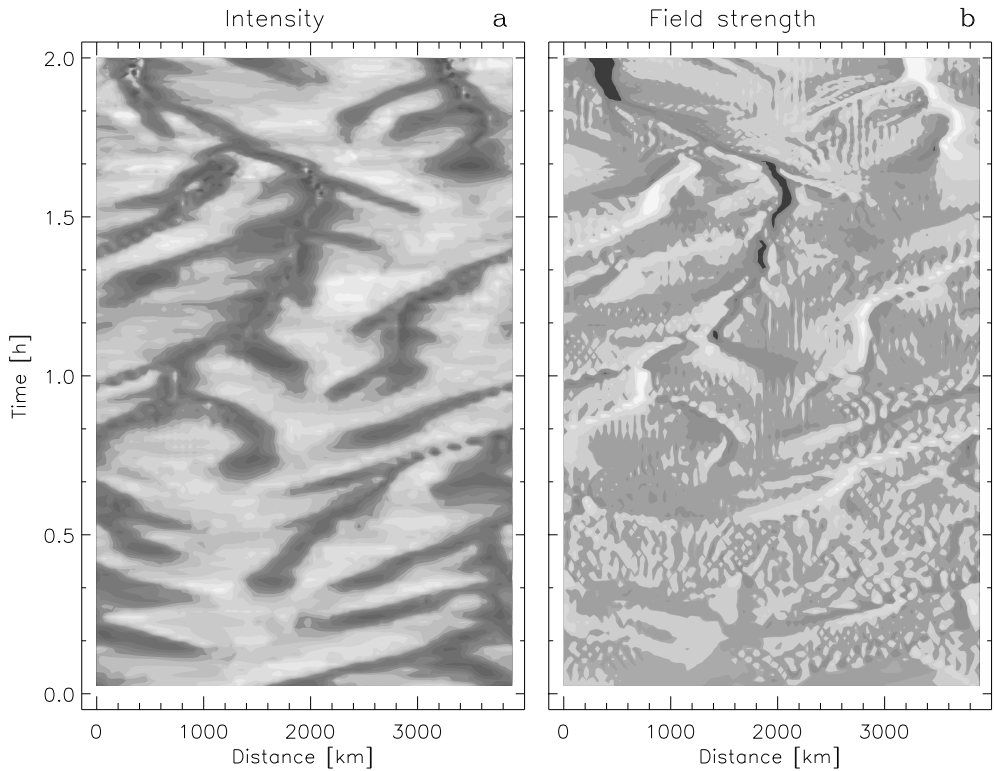}

   \caption[]{
Spatial and temporal evolution of the emerging monochromatic intensity
$I_c/{<}I_c{>}$ at $\lambda$ 500~nm (a) and the magnetic field intensity $B$ at the
$\log \tau_R = 0$ level (b) in the 2-D MHD simulation. The intensity
$I_c/{<}I_c{>}$ ranges from 0.017 to 1.78 (on a linear scale). The field intensity
$B$ ranges from -286.7 to +254.0~mT. Dark hatching) negative polarity field, light
hatching) positive polarity field.
 }
      \label{Fig:Fig3}
\end{figure}

Figure~\ref{Fig:Fig2} displays the average value of $B$ in the simulation region as
a function of magnetohydrodynamic (real solar) time. The simulation may be
arbitrarily divided into three stages -- the initial period (to 20 min), which is
mainly controlled by the initial conditions adopted for the magnetic field, the
transition period, when a mutual rearrangement of the thermal convection and the
magnetic field occurs, and finally, the period of self-consistent solutions.

Figure~\ref{Fig:Fig3} demonstrates the evolution of the magnetic field and the
brightness field (visible granulation pattern at a wavelength of 500~nm). The
intergranular lanes in Fig.~\ref{Fig:Fig3}a are shown by dark hatching and granules
are shown by light hatching. The granules and intergranular lanes were
distinguished for every individual model with respect to the mean intensity level
at A 500~nm. The areas with intensities above this level were recognized as
granules and below this level as intergranular lanes.

The magnetic field for the continuum formation level ($\log \tau_R = 0$, $\tau_R$
being the Rosseland optical depth) is shown in Fig.~\ref{Fig:Fig3}b with the field
polarities hatched differently. In the course of the initial 20 min the
concentration and intensification of the magnetic field occur in the vicinity of
downflows ($x = 3600$--3800~km). Then a partial reconnection of field lines takes
place (the reason is, in part, that the initial field was taken as bipolar), and
local magnetic field concentrations dissipate. Over the period from the 20th to
35th minute there are no marked structures in the magnetic field. A clear-cut
structuring in the field concentration is observed afterwards: the field becomes
stronger in intergranular lanes. The onset of concentration (formation of new thin
magnetic elements) coincides in time and space with the moment when granules break
up, and the field separates polarity at that time as well. The intergranular lanes
become brighter when the intensity of magnetic features exceeds 100~mT.

The fragmentation of convective flows is due to the action of surface layers, which
serve as a thermal boundary \cite{Rast93}. For example, the Rosseland absorption
coefficient decreases by the order of 1000 and the heat capacity by the order of 10
in hot upflows near the visible surface in a height interval as small as 100 km
\cite{Gadun98}. As a result, the hot thermal flows are intensely cooled (due mainly
to radiation losses) and become unstable. As shown in
\cite{Gadun95a,Gadun96,Ploner}, the mechanism through which the flows lose their
stability is determined by their horizontal scale. The flows related to the
granules up to $l_g = 1000$--1400~km in size dissipate in the main, while larger
flows break up. It was found in \cite{Gadun98} from 2-D~HD models that of crucial
importance for small-scale flows ($l_g \leq 600$~km) is the thermal decay due to
radiation losses predominantly in the horizontal direction. Such flows cool rapidly
and dissipate.

\begin{figure}
   \centering
   \includegraphics[width=14.5cm]{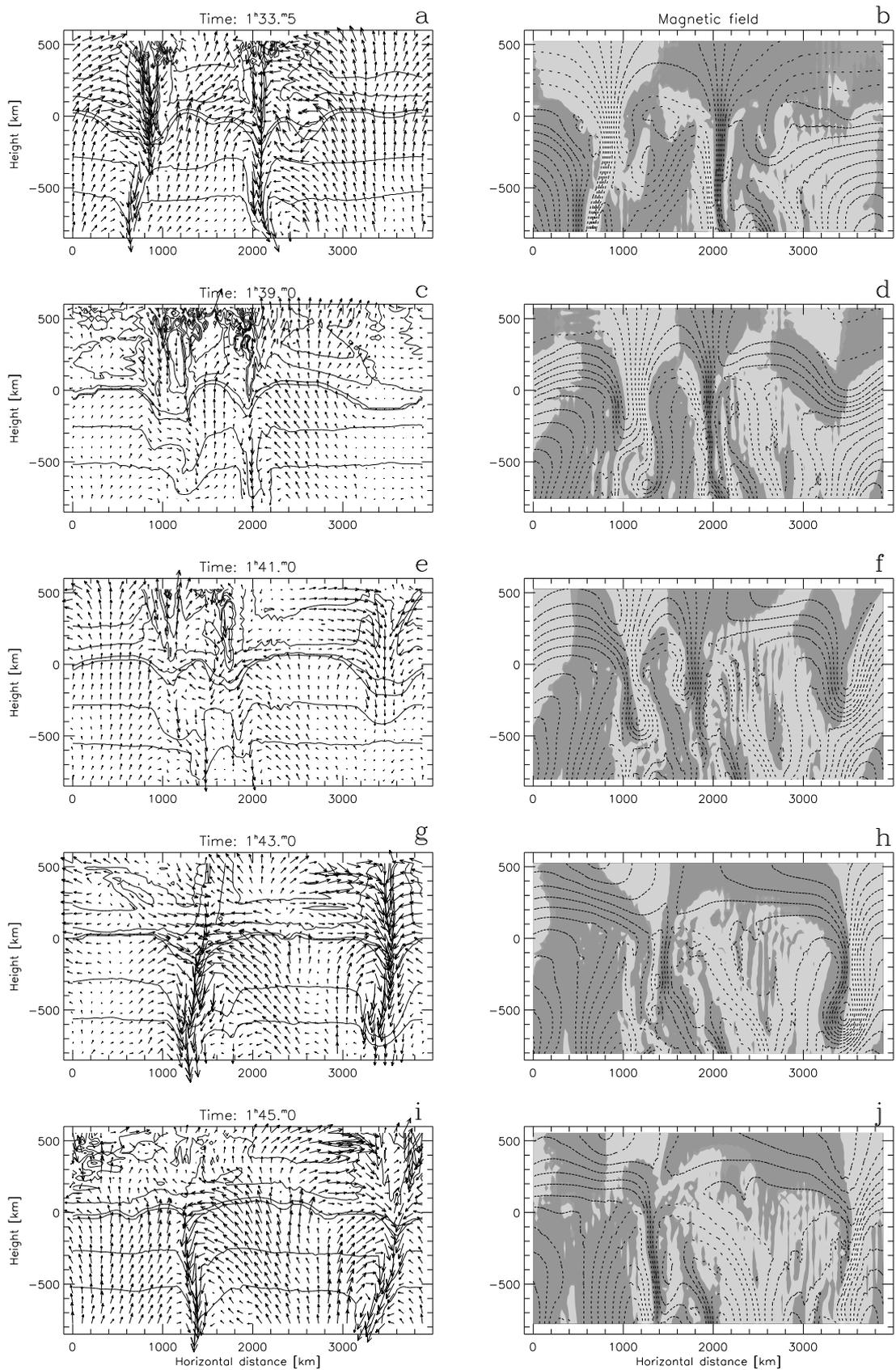}
   \caption[]{
Left panels) snapshots of velocity field and temperature. Horizontal lines are
4000, 5000, 6000, 7000, 12000, and 13000 K isotherms (from the top down). Right
panels) snapshots of magnetic field lines (dotted) and field polarity  with dark
hatching for negative polarity field and light hatching for positive polarity
field.
 }
       \label{Fig:Fig4}
\end{figure}

\begin{figure}
   \centering
 \includegraphics[width=14.5cm]{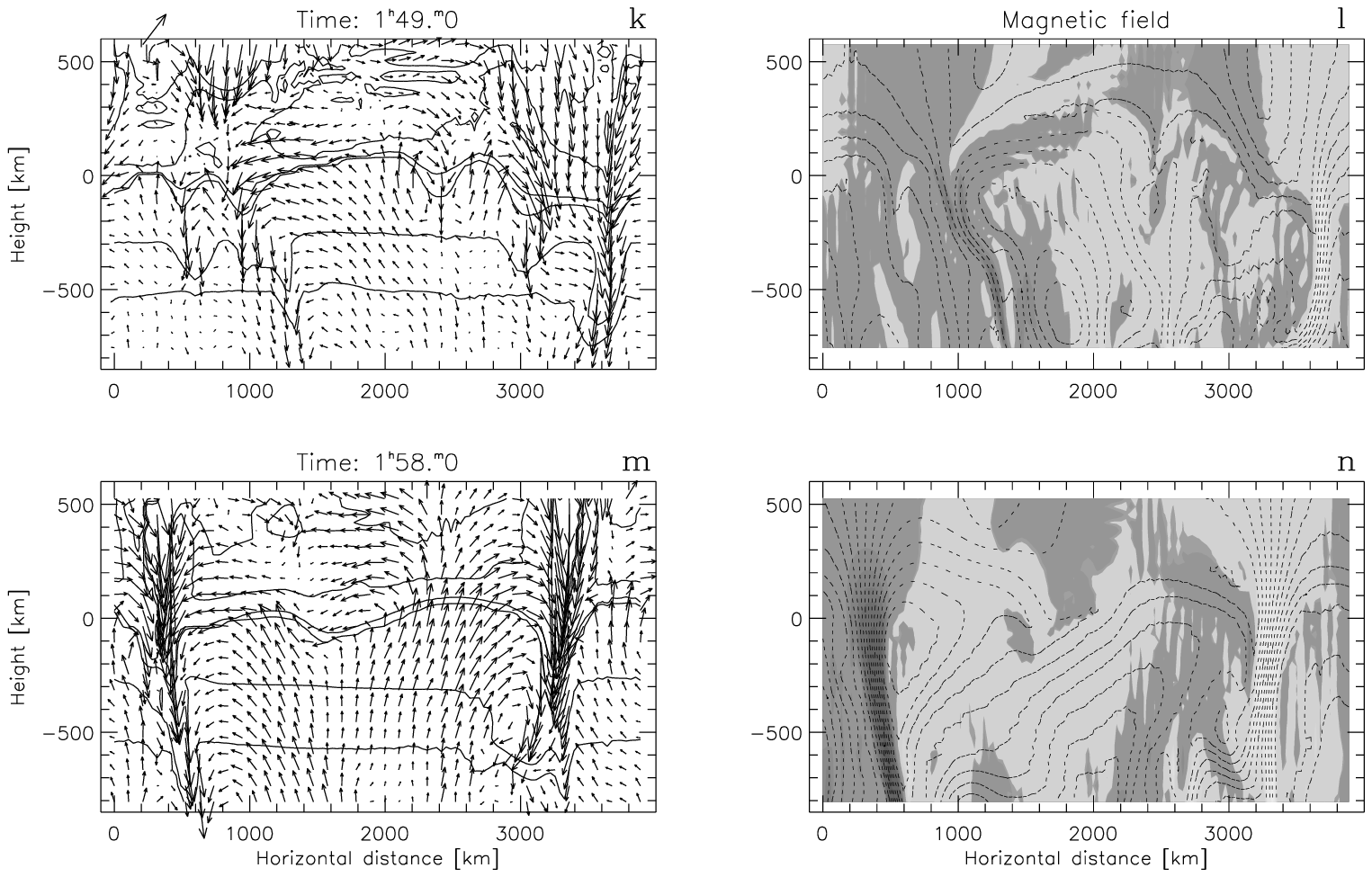}
   \hfill
   \parbox[b]{16.cm}{ \vspace{0.5cm}
 {Figure 4 (continued):
 Snapshots of velocity field and temperature as well as magnetic field lines and
field polarity. Designations are the same.  }

      }
\end{figure}

Large-scale convective upflows with $l_g \geq 1000$~km are highly adiabatic, and
the major cause of their instability is the excess pressure which develops inside
them. This pressure excess is a fundamental property of the thermal in convection
in stratified media (this was found even with the simplest polytropic models
\cite{Bray}). Its relative magnitude is especially large near the surface (the
thermal boundary), and this gives rise to the buoyancy inversion. It is the
locality where the adiabatically moving matter penetrates into the photosphere and
is slowed down through the radiation cooling. In addition, large horizontal
dimensions of convection cells favor the formation of a highly dense passive cloud
of matter which hangs over the central parts of convection cells. The excessive
pressure forces weaken the central convective flow and tend to redistribute the
convective energy flux in the horizontal direction from central regions to
periphery. Thereafter the weakened central fragment of the large-scale thermal flow
cannot provide the balance of forces to support stability near the surface, and a
stream of matter directed from the surface downwards is formed from the dense
cloud; the fragmentation thus comes to an end \cite{Gadun98}.

It was shown in \cite{Gadun98} that the flows of intermediate horizontal dimensions
($l_g = 600$--1000~km) experience the effect of both mechanisms to the least
extent. These theoretical results found an observational support
\cite{Karpinskii,Kawaguchi}.

The fragmentation of the thermal convective flows in the presence of magnetic
fields maintains the formation of new small-scale magnetic elements. It this case
the streams of cool matter newly formed near the surface entrap horizontal magnetic
field lines, carry them downwards, changing the horizontal orientation of the field
to the vertical one, and intensify the field by extending the field lines. The
photospheric horizontal magnetic field is quite week, and it meets the freezing-in
condition. The field lines are not completely reconnected in this case, since the
velocity field is not completely symmetrical in the convective cell which undergoes
fragmentation and the cell itself displays clear-cut polarities, with the field
intensity growing with depth.

Figure~\ref{Fig:Fig4} illustrate the mechanism described above. The time in this
figures corresponds to the time scale in Fig.~\ref{Fig:Fig3}, so that the
magnetogranulation evolution can be followed in parallel in both figures. The first
fragment represents the situation obtained in the calculation region after
$1^h33.5^m$ of modeling. Two flux tubes of opposite polarities are concentrated in
intergranular lanes. In the right part of the region ($x > 2200$~km) a
well-developed convective flow (a granule) is located and suppression of the
central fragment has already begun in the lower part of this flow.

A characteristic feature at the moment $1^h39^m$ is the onset of fragmentation in
the large-scale flow: the surface level between $x \approx 3300$~km and $x \approx
3700$~km is sagged, and the horizontal magnetic field lines begin to sag as well.

The fragmentation progresses further from $1^h41^m$ to $1^h43^m$: an active
downflow develops due to the convective instability. The field lines are entrained
by it. No complete reconnection of the entrained field lines is observed, as the
convective cell itself has clear-cut polarities in the area where the downflow is
passing and the field intensity grows with depth inside the sell in accordance with
the equipartition condition. The field of one polarity only is intensified (the
positive polarity in this case), and the field of the other polarity weakens
($1^h45^m$).

Finally, in the time interval from $1^h45^m$ to $1^h58^m$ well-developed flux tubes
are observed: one tube ($x \approx 3200$--3400~km) was formed by the above
mechanism, and the second tube ($x \approx 400$--600~km) appeared due to the
kinematic effect. However, the superadiabatic (or thermal) mechanism plays the
major role in the field intensification in both cases; it is associated with the
superconvective instability of downflows in the presence of strong magnetic fields.

The figures demonstrate not only the formation of flux tubes but their dissipation
as well -- the tube found between $x = 600$~km and $x = 800$~km at $1^h33.5^m$
dissipated completely by the moment $1^h43^m$, and another compact magnetic
configuration, between 2000 and 2200~km, weakens significantly and disappears as a
bright feature in an intergranular region.

Compact fields dissipate in our models due to two mechanisms: the internal
instability of such intensive small-scale features and their weakening caused by
reconnection of field lines. The internal instability of flux tubes was studied in
detail in [13]: it develops in intensive magnetic structures of a larger scale as a
result of potent convective instability -- this instability gives rise to an
intense evaporation of the matter in the subphotospheric layers inside the tube.
Under the conditions governing in a gravitationally stratified medium this
inevitably leads to the expansion of matter along field lines from deeper layers
and the "reversion" of the superadiabatic effect -- downflows are replaced by
upflows inside the tube. The fields become less compact, and the tube breaks down.

The other mechanism is clearly recognized in our simulation. It, in essence, is
this: the tube and the adjacent regions of convective cells may be of different
polarities, and the kinematic effect, which steadily squeezes the field out of
cells, weakens the tube in the intergranular region owing to the reconnection of
field lines.

An interesting effect is produced by the common operation of the kinematic
mechanism and the thermal one in the concentration and intensification of the
magnetic field -- newly formed compact structures with strong magnetic fields and
areas with relatively weak fields of opposite polarity may be next to one another.
This effect can be seen in Figs~\ref{Fig:Fig3} and ~\ref{Fig:Fig4}, and it is
confirmed by observations of magnetic regions with high spatial resolution
\cite{Koutchmy1,Koutchmy2}.

Thus, the role of granulation in the formation of small-scale magnetic structures
should be revised. While it was originally assumed that it only concentrates the
global large-scale magnetic field in bundles at cell boundaries, our calculations
suggest that it is capable of forming thin magnetic tubes from the weak
photospheric horizontal field. The existence of this field is confirmed by
observations. In his controversial paper \cite{Koutchmy1} Sch\"{u}ssler gives the
observed distribution of the magnetic field inclination as a function of field
intensity. This distribution suggests that weak fields are predominantly horizontal
in the photospheric layers, and so the mechanism described above is quite
realistic.

\begin{figure}
   \centering
   \includegraphics[width=6.cm]{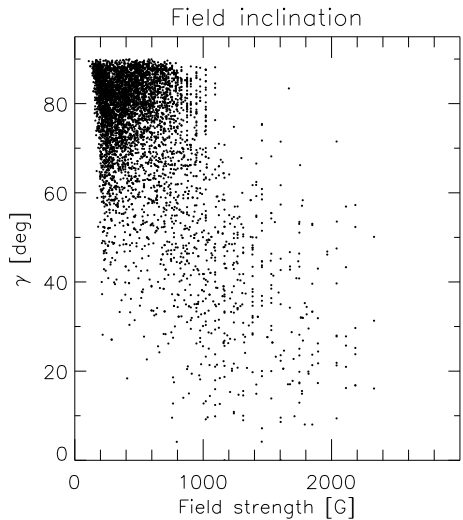}
 \hfill
\parbox[b]{7.cm}{ \vspace{0.0cm}
   \caption[]{
The magnetic field inclination $\gamma$ vs. the field intensity $B$.
 }\label{Fig:Fig6}
}

\end{figure}

We obtained a similar distribution  from our models by simulating spectral
observations during the last 25 min of the MHD modeling (Fig.~\ref{Fig:Fig6}).
First we calculated the Stokes profiles for the IR Fe I line 1564.8~nm (it is often
used in observation programs). Then we applied the well-known Stokes diagnostics
techniques to the theoretical spectral ``scans'' in order to determine the
inclination of the magnetic field to the vertical, $\gamma$, and the field
intensity $B$. The inclination angle was found from the relation $\tan \gamma =
\sqrt{Q}{/}V$ (this is an approximate relation, and it is evident from the known
expressions for the amplitudes of the $\sigma$-components of the V, Q, and U Stokes
profiles), and the field intensity was estimated from the distance between the
peaks in the red and blue wings of the V profile. In view of a lower spatial
resolution of the observations, the agreement between the observed and theoretical
distributions is quite good. This is one more argument for revising the concept f
the formation of small-scale magnetic elements.

\section{Conclusion
}

A direct numerical simulation of the magnetoconvection on the granulation scales
suggests that the role of granulation in the formation of small-scale magnetic
structures should be revised. The granulation can concentrate the magnetic field in
bundles on the boundaries of convective cells and enhance the field intensity in
them, but it can also form small-scale flux tubes from weak horizontal fields,
which may be, in principle, of local nature.

{\bf Acknowledgements.} This study was partially financed by the Swiss National
Science Foundation (Grant No. 7UKPJ 48440).



\end{document}